\documentclass[aps,pra,showpacs,superscriptaddress,twocolumn,a4paper]{revtex4}

\usepackage{amsmath}
\usepackage{amsfonts}
\usepackage{amssymb}
\usepackage{graphicx}
\usepackage{color}
\usepackage[colorlinks=true,citecolor=red,linkcolor=blue]{hyperref}

\newcommand{\ket}[1]{\left| #1 \right\rangle}
\newcommand{\bra}[1]{\left\langle #1\right |}

\def\mathbi#1{\textbf{\em #1}}

\begin{document}

\title{Non-Hermitian wave packet approximation of Bloch optical equations}

\author{Eric Charron}
\affiliation{Universit\'e Paris-Sud, Institut des Sciences Mol\'eculaires d'Orsay,\\ ISMO, CNRS, F-91405 Orsay, France}

\author{Maxim Sukharev}
\affiliation{Department of Applied Sciences and Mathematics, Arizona State University, Mesa, Arizona 85212, USA}

\date{\today}

\begin{abstract}

We introduce a non-Hermitian approximation of Bloch optical equations. This approximation provides a complete description of the excitation,
relaxation and decoherence dynamics of ensembles of coupled quantum systems in weak laser fields, taking into account collective effects and
dephasing. In the proposed method one propagates the wave function of the system instead of a complete density matrix. Relaxation and dephasing
are taken into account via automatically-adjusted time-dependent gain and decay rates. As an application, we compute the numerical wave packet solution
of a time-dependent non-Hermitian Schr\"odinger equation describing the interaction of electromagnetic radiation with a quantum nano-structure
and compare the calculated transmission, reflection, and absorption spectra with those obtained from the numerical solution of the Liouville-
von-Neumann equation. It is shown that the proposed wave packet scheme is significantly faster than the propagation of the full density matrix
while maintaining small error. We provide the key ingredients for easy-to-use implementation of the proposed scheme and identify the limits and
error scaling of this approximation.

\end{abstract}


\maketitle

\section{Introduction}

Optics of nanoscale materials has attracted considerable attention in the past several years \cite{Maier:2005zp,Lal:2007xy,
Ozbay:2006pv,Berini:09} due to various important applications \cite{Stockman:2011bw}. Exploring electrodynamics of near-fields associated
with subwavelength systems, researchers are now truly dwelling into nanoscale \cite{Maier:2001gz,Garcia-de-Abajo:2007py,
Ebbesen:2008ab,Schuller:2010zi,Stockman:2011bw}. Owing to both new materials processing techniques \cite{Hutter:2004ae} and
continuous progress in laser physics \cite{Slusher:1999vd} the research in nano-optics is currently transitioning from linear
systems, where materials and their relative arrangement control optical properties \cite{Barnes:2007lr}, to the nonlinear regime
\cite{Kroo:2008pi}. The latter expands optical control capabilities far beyond conventional linear optics as in the case of active
plasmonic materials \cite{Wuestner:2011xx}, for instance, combining highly localized electromagnetic (EM) radiation driven by
surface plasmon-polaritons (SPP) with non-linear materials~\cite{Krasavin:2006qj}.
Yet another promising research direction, namely optics of highly coupled exciton-polariton systems, is emerging
\cite{Bellessa:2004ec,Chang:2006mh}. It basically reincarnates a part of research in semiconductors \cite{Agranovich:2005mr,
Khitrova:2006jq}, bringing it to nanoscale via deposition of ensembles of quantum emitters (molecules \cite{Dintinger:2005ll,
Ebbesen:2009ua,Lekeufack:2010mm,Berrier:2011fe}, quantum dots \cite{Park:2007tk,Komarala:2008fp,Andersen:2011wt,
doi:10.1021/nl200052j}) directly on to plasmonic materials.

Even in the linear regime, when the external EM radiation
is not significantly exciting the quantum sub-system, SPP near-fields can be strongly coupled to quantum emitters. This manifests
itself as a Rabi splitting widely observed in transmission experiments \cite{Gomez:2010xi}. Moreover, a new phenomenon, namely
collective molecular modes driven by SPP near-fields, has been observed \cite{Ebbesen:2009ua} and recently explained \cite{Salomon:2012xk}.
It was also shown that nanoscale clusters comprised of optically coupled quantum emitters exhibit collective scattering and absorption
\cite{Sukharev:2011ib} similar to Dicke superradiance \cite{DickeUFNreview}. It is hence important to be able to account for
collective effects in a self-consistent manner. We also note that a series of works by Neuhauser \textit{et al.} \cite{gupta2001rabi,Lopata:2007qe,
Lopata:2009ao,Lopata:2009jq} clearly demonstrated that the presence of a single molecule nearby a plasmonic material can
significantly alter the scattering spectra.

In many applications (such as optics of molecular layers coupled to plasmonic materials \cite{Dintinger:2005ll,Ebbesen:2009ua,
Lekeufack:2010mm,Berrier:2011fe,Salomon:2012xk}, for instance) self-consistent modeling relies heavily on the numerical integration of
the corresponding Maxwell-Bloch equations, assuming that static emitter-emitter interactions can be neglected, which is true for systems
at relatively low densities. Such an approximation results in expressing the local polarization in terms of a product of the local
density of quantum emitters and the local averaged single emitter's dipole moment \cite{Allen:1975ij}. One of the first efficient
numerical schemes for simulations of nonlinear optical phenomena of quantum media driven by external classical EM radiation was
proposed by Ziolkowski et al. \cite{Ziolkowski:1995qd}. Using a one-dimensional example of ensembles of two-level atoms it was
shown that the corresponding Maxwell-Bloch equations can be successfully integrated using an iterative scheme based on the
predictor-corrector method (\textit{strongly coupled method}). Later on this approach has been extended to two-
\cite{Slavcheva:2002id} and three-dimensional systems \cite{Fratalocchi:2008fs}. Although such a scheme accurately captures
the system's dynamics, it can become extensively slow for multidimensional systems \cite{Sukharev:2011ib}. Moreover this method is limited to
two-level systems only. A more efficient technique based on the decoupling of Bloch equations from the Ampere law was proposed in
2003 by Bid\'egaray \cite{Bidegaray:2003op}. This latter method, usually referred to as a \textit{weakly coupled method},
noticeably improves the efficiency of the numerical integration of Maxwell-Bloch equations, allowing to consider multilevel
quantum media \cite{Sukharev:2011ib}.

The approach proposed in the present paper is based on this weakly coupled method \cite{Sukharev:2011ib}. It further improves the numerical
efficiency of the Maxwell-Bloch integrator for ensembles of multilevel quantum emitters. By incorporating a new non-Hermitian wave packet
propagation technique into the weakly coupled method, we demonstrate that our approach can be successfully applied to ensembles of multilevel
atoms and diatomic molecules. We show with this method that it is sufficient to propagate a single wave function instead of the complete
density matrix. Relaxation and dephasing are taken into account via empirical gain and decay rates whose time-dependence is automatically
adapted for an optimal description of dephasing processes.

The paper is organized as follows. We first introduce our model in section \ref{sec:tls}, using, as an example, an ensemble of two-level atoms.
The non-Hermitian wave packet approximation is then described in section \ref{sec:nhtlwpa}. Applications of the proposed method to the case
of a nano-layer comprised of two-level atoms are discussed in section \ref{sec:aunla}. We then generalize our method to the case of interacting
multi-level emitters in section \ref{sec:multilevels}. Finally, using this generalized approach, we calculate in section \ref{sec:molecular_layer}
the optical properties of a nano-layer of coupled molecules, taking into account both the vibrational and rotational degrees of freedom, and we
reveal several interesting features in the absorption spectra. Last section \ref{sec:summary} summarizes our work.

\section{Theoretical model and applications}

As pointed out in the previous section, the importance of collective effects manifested in recent experiments and theoretical papers
calls for the development of self-consistent models capable of taking into account mutual EM interaction in ensembles of quantum emitters.
While direct numerical integration of corresponding Maxwell-Bloch equations can be based on either a strong coupled method
\cite{Ziolkowski:1995qd} or a weakly coupled one \cite{Bidegaray:2003op}, for multi-level systems in two- and especially
three-dimensions such a brute-force approach becomes numerically very expensive, both in terms of computation (CPU) times and
memory requirements. Indeed, the main disadvantage of such approaches is that both the CPU times and memory requirements
scale generally at least as $N_h^2$, where $N_h$ is the dimension of the Hilbert space of the system. This unfavorable scaling is
directly related to the size of the reduced density matrix used to describe the system and to the associated number of time-dependent
equations one has to solve to follow the system's dynamics. By contrast, one has only to solve a reduced set of $N_h$ time-dependent
equations when the system can be described by a single wave function. It would therefore be extremely useful to be able to derive a
``Schr\"odinger-type'' approximation which could describe on an equal footing the field-induced coherent dynamics of a multi-level
quantum system and the associated relaxation and decoherence processes. This type of idea is not entirely new. For small excitations,
the time-dependent density functional theory \cite{PhysRevLett.52.997} allows for instance, in its real-time version, the extraction of density matrix
dephasing without evolving the full density matrix. This goal has also been achieved in the past using different approaches,
the three main contributions being the stochastic Schr\"odinger method used in conjunction with a Monte Carlo integrator \cite{PhysRevLett.68.580,
PhysRevA.45.4879, PhysRevA.46.4363, PhysRevA.46.4382, 0305-4470-25-21-023, makarov:10126}, the Gadzuk jumping wave packet scheme \cite{Gadzuk1990317,
Finger1997291} and the variational wave packet method \cite{gerdts:3017, Pesce1998383}. These methods require the propagation of $N_f$ different wave
functions and are therefore mainly attractive when $N_f \ll N_h$.

The approach we propose here is different since it is based on perturbation theory. It also significantly speeds up calculations
since it only requires the propagation of a \textit{single} wave function under the action of an easy-to-implement time-dependent
effective Hamiltonian in order to reproduce accurately the full dynamics. We demonstrate the efficiency and accuracy of our method
using, first, ensembles of two-level atoms, and then, interacting diatomic molecules including both the vibrational and rotational
degrees of freedom. We also discuss the conditions under which our method is no longer valid. All results are compared with data
obtained via direct integration of Maxwell-Bloch equations using a weakly coupled method \cite{Sukharev:2011ib}.

\subsection{Atomic two-level system}
\label{sec:tls}

Let us first consider a two-level quantum system interacting with EM radiation. We label the two levels as $\ket{0}$ and $\ket{1}$,
with associated energy eigenvalues $\hbar\omega_0$ and $\hbar\omega_1$, respectively. The corresponding density matrix
$\hat\rho(t)$ satisfies the dissipative Liouville-von Neumann equation \cite{Blum:1996kn}
\begin{equation}
 \label{eq:Liouville}
 i\hbar\frac{\partial\hat\rho}{\partial t} = [ \hat H , \hat \rho ] -i\hbar \hat\Gamma \hat\rho
\end{equation}
where $\hat H = \hat H_0 + \hat V_i(t)$ is the total Hamiltonian and $\hat\Gamma$ is a superoperator, taken in the Lindblad form \cite{Breuer02},
describing relaxation and dephasing processes under Markov approximation. The field free Hamiltonian reads
\begin{equation}
 \label{eq:H0}
 \hat H_0 = \hbar\omega_0 \ket{0}\!\!\bra{0} + \hbar\omega_1 \ket{1}\!\!\bra{1}.
\end{equation}
The interaction of the two-level system with EM radiation is taken in the form
\begin{equation}
 \label{eq:Vi}
 \hat V_i(t) = \hbar\Omega(t) \big( \ket{1}\!\!\bra{0} + \ket{0}\!\!\bra{1} \big).
\end{equation}
$\Omega(t)$ denotes here the instantaneous Rabi frequency associated with the coupling between the quantum system and an external
field. In Eq.\,(\ref{eq:Liouville}), the non-diagonal elements of the operator $\hat\Gamma$ include a pure dephasing rate $\gamma^*$,
and the diagonal elements of this operator consist of the radiationless decay rate $\Gamma$ of the excited state. Equations
(\ref{eq:Liouville})-(\ref{eq:Vi}) lead to the well-known Bloch optical equations \cite{Allen:1975ij} describing the quantum dynamics
of a coupled two-level system
\begin{subequations}
 \begin{eqnarray}
  \dot{\rho}_{00} & = & i\Omega(t)\left(\rho_{01}-\rho_{10}\right) + \Gamma\rho_{11} \label{rho00}\\
  \dot{\rho}_{01} & = & i\Omega(t)\left(\rho_{00}-\rho_{11}\right) + \left[i\omega_{\!B} - \frac{2\gamma^*+\Gamma}{2}\right]\rho_{01} \label{rho01}\\
  \dot{\rho}_{10} & = & i\Omega(t)\left(\rho_{11}-\rho_{00}\right) - \left[i\omega_{\!B} + \frac{2\gamma^*+\Gamma}{2}\right]\rho_{10} \label{rho10}\\
  \dot{\rho}_{11} & = & i\Omega(t)\left(\rho_{10}-\rho_{01}\right) - \Gamma\rho_{11} \label{rho11}
 \end{eqnarray}
\end{subequations}
where $\omega_{\!B}=\omega_1-\omega_0$ is the Bohr transition frequency and where the dot denotes the time derivative.

We assume in the following that the system is initially in the ground state $\ket{0}$, and we will show that, under some assumptions, the
subsequent induced excitation and relaxation dynamics can be accurately described by a non-Hermitian wave packet approximation.

\subsection{Non-Hermitian two-level wave packet approximation}
\label{sec:nhtlwpa}

Within the aforementioned simplified model, the system's wave packet $\ket{\Psi(t)}$ can be expanded as
\begin{equation}
 \ket{\Psi(t)} = c_0(t) \ket{0} + c_1(t) \ket{1}. \label{expansion}
\end{equation}
Let us now assume that the ground and excited states energies include an imaginary part that we denote as $+\hbar\gamma_0/2$ and
$-\hbar\gamma_1/2$, respectively. Inserting expansion\,(\ref{expansion}) into the time-dependent Schr\"odinger equation
\begin{equation}
 i\hbar\frac{\partial}{\partial t}\ket{\Psi(t)} = \hat{H}\ket{\Psi(t)} \label{Schrodinger}
\end{equation}
and projecting it onto states $\ket{0}$ and $\ket{1}$ yields the following set of coupled equations for the coefficients $c_n(t)$
\begin{subequations}
 \begin{eqnarray}
  i\,\dot{c}_0 & = & \left( \omega_0 + i\frac{\gamma_0}{2} \right) c_0 +                  \Omega(t)\,                  c_1    \label{c0} \\
  i\,\dot{c}_1 & = &                    \Omega(t)\,                c_0 + \left( \omega_1 - i\frac{\gamma_1}{2} \right) c_1\,. \label{c1}
 \end{eqnarray}
\end{subequations}

We will now derive the differential equations describing the temporal dynamics of the products $\rho^s_{ij} = c_i\,c_j^*$, where the
subscript $s$ corresponds to this simplified non-Hermitian Schr\"odinger-type model. Our goal is to obtain the gain and decay rates $\gamma_0$
and $\gamma_1$ which will allow for an approximate description of the system's excitation and relaxation dynamics. For the evolution of the
populations $\rho^s_{00}(t)$ and $\rho^s_{11}(t)$, one gets
\begin{subequations}
 \begin{eqnarray}
  \dot{\rho}^s_{00} & = & i \Omega(t)\left(\rho^s_{01}-\rho^s_{10}\right) + \gamma_0\,\rho^s_{00}\\
  \dot{\rho}^s_{11} & = & i \Omega(t)\left(\rho^s_{10}-\rho^s_{01}\right) - \gamma_1\,\rho^s_{11}\,.
 \end{eqnarray}
\end{subequations}
The conservation of the total norm
\begin{equation}
 \dot{\rho}^s_{00}(t) + \dot{\rho}^s_{11}(t) = 0
\end{equation}
thus results in
\begin{equation}
 \gamma_0\,\rho^s_{00}(t) = \gamma_1\,\rho^s_{11}(t) \label{norme}
\end{equation}
It therefore appears that, since the populations $\rho^s_{00}(t)$ and $\rho^s_{11}(t)$ are generally time dependent, \textit{at least one
of the two rates} $\gamma_0$ \textit{or} $\gamma_1$, which are not yet fully defined, \textit{must be taken as a time-dependent function}.

Taking Eq.\,(\ref{norme}) into account, one finally obtains the following set of Bloch equations for the approximate density matrix $\rho^s(t)$
\begin{subequations}
 \begin{eqnarray}
 \dot{\rho}^s_{00} & = & i\Omega(t)\left(\rho^s_{01}-\rho^s_{10}\right) + \gamma_1\,\rho^s_{11} \label{rhos00}\\
 \dot{\rho}^s_{01} & = & i\Omega(t)(\rho^s_{00}-\rho^s_{11}) + \!\left[ i\omega_{\!B} - \frac{\gamma_1-\gamma_0}{2} \right]\!\rho^s_{01} \label{rhos01}\\
 \dot{\rho}^s_{10} & = & i\Omega(t)(\rho^s_{11}-\rho^s_{00}) - \!\left[ i\omega_{\!B} + \frac{\gamma_1-\gamma_0}{2} \right]\!\rho^s_{10} \label{rhos10}\\
 \dot{\rho}^s_{11} & = & i\Omega(t)\left(\rho^s_{10}-\rho^s_{01}\right) - \gamma_1\,\rho^s_{11} \label{rhos11}
\end{eqnarray}
\end{subequations}
to be compared with the exact equations\,(\ref{rho00})-(\ref{rho11}).

Two obvious choices can then be made for the empirical gain and decay parameters $\gamma_0$ and $\gamma_1$:
\begin{itemize}
 \item The first choice $\gamma_1=\Gamma$ allows to reproduce correctly the equations\,(\ref{rho00}) and\,(\ref{rho11}) describing the
       populations at the cost of degrading the description of the coherences $\rho_{01}(t)$ and $\rho_{10}(t)$.
 \item The second choice, $\gamma_1-\gamma_0=2\gamma^*+\Gamma$, allows to reproduce correctly the equations\,(\ref{rho01}) and\,(\ref{rho10})
       describing the coherences, at the cost of degrading the description of the populations.
\end{itemize}
The optimal choice for applications in linear optics of nano-materials should be as follows: in weak fields, the variation of the populations, as
described by perturbation theory, is a second order term with respect to the coupling amplitude while the variation of the coherences is a first
order term. For a correct description of the quantum dynamics in weak fields, it is important to describe first order terms accurately. Therefore
we proceed with the second choice
\begin{equation}
 \gamma_1-\gamma_0 = 2\gamma^*+\Gamma. \label{coh}
\end{equation}

From Eqs.(\ref{norme}) and\,(\ref{coh}) we obtain a set of empirical time dependent gain and decay rates $\gamma_0(t)$ and $\gamma_1(t)$ that we
can insert in the Schr\"odinger-type approximation
\begin{subequations}
 \begin{eqnarray}
  \gamma_0(t) & = & \frac{(2\gamma^*+\Gamma)|c_1(t)|^2}{|c_0(t)|^2-|c_1(t)|^2} \label{g0}\\
  \gamma_1(t) & = & \frac{(2\gamma^*+\Gamma)|c_0(t)|^2}{|c_0(t)|^2-|c_1(t)|^2} \label{g1}
 \end{eqnarray}
\end{subequations}
We finally obtain the following set of time-dependent
coupled equations
\begin{subequations}
 \begin{eqnarray}
  i\,\dot{c}_0 & = & \left( \omega_0 + i\frac{(\gamma^*+\Gamma/2)|c_1|^2}{|c_0|^2-|c_1|^2} \right) c_0 + \Omega(t)\,c_1 \label{c0bis} \\
  i\,\dot{c}_1 & = & \Omega(t)\,c_0 + \left( \omega_1 - i\frac{(\gamma^*+\Gamma/2)|c_0|^2}{|c_0|^2-|c_1|^2} \right) c_1 \label{c1bis}
 \end{eqnarray}
\end{subequations}
This system is solved numerically using a fourth order Runge-Kutta algorithm \cite{NR2007}. Eqs. (\ref{c0bis}),  (\ref{c1bis}) include two
non-Hermitian terms which can (as we will illustrate below) accurately reproduce the dissipative dynamics in weak fields, \textit{i.e.} for
$|c_1|^2 \ll |c_0|^2 \approx 1$. Indeed, as one can easily show that, in this limit, Eqs.\,(\ref{rhos00})-(\ref{rhos11}) are strictly equivalent to
the exact Bloch Eqs.\,(\ref{rho00})-(\ref{rho11}).

At first sight, it might seem that the proposed formalism rather relies on a mathematical trick. It does however have physical insights.
Indeed, the time-dependent gain and decay rates, in some cases, have a deep physical meaning, as it was
shown in Ref. \cite{PhysRevLett.101.080402}, for instance, where the non-Hermitian formalism was applied to study EM wave propagation
in so-called PT-symmetric waveguides. In this study, it was shown that the gain and loss coefficients could be used to control
the beat length parameter which describes waveguides. In our approach $\gamma_0(t)$ and $\gamma_1(t)$ are gain and decay rates inherent
to the studied quantum system since they only depend on the decay and dephasing rates $\Gamma$ and $\gamma^*$ of this system and on the
relative population of the quantum states involved. In the past, such type of non-Hermitian approaches have been proven to be very
powerful tools in many branches of physics, including resonant phenomena, quantum mechanics, optics, and quantum field theory. A
recent comprehensive survey of various applications of the non-Hermitian approach can be found in Ref. \cite{moiseyev2011non}.

\subsection{Application to a uniform nano-layer of atoms}
\label{sec:aunla}

While our approximation is equally applicable to three-dimensional systems, we have chosen, for the sake of simplicity, to test it on a
simplified one-dimensional system consisting of a uniform infinite layer of atoms whose thickness $\Delta z$ lies in the range of a few
hundred nanometers. An incident radiation field propagating in the positive $z$-direction is represented by a transverse-electric mode
with respect to the propagation axis. It is characterized by a single in-plane electric field component $E_x(z,t)$ and a single in-plane
magnetic field component $H_y(z,t)$. To account for the symmetry of the atomic polarization response, the atoms in the layer are described
as two-level systems with the following states: an $s$-type ground state and a $p_x$-type excited state. This model is a one-dimensional
simplification of a more general approach used in Ref. \cite{Sukharev:2011ib} and we refer the reader to the body of this paper for details.

The time-domain Maxwell's equations for the dynamics of the electromagnetic fields
\begin{subequations}
 \begin{eqnarray}
  \mu_0\,\partial_t H_y      & = & -\partial_z E_x \label{Faraday} \\
  \epsilon_0\,\partial_t E_x & = & -\partial_z H_y -\partial_t P_x \label{Ampere}
 \end{eqnarray}
\end{subequations}
are solved using a generalized finite-difference time-domain technique where both the electric and magnetic fields are propagated in discretized
time and space \cite{Taflove:2005jj}. In these equations, $\mu_0$ and $\epsilon_0$ denote the magnetic permeability and dielectric permittivity
of free space. The macroscopic polarization of the atomic system
\begin{equation}
 P_x(z,t)= n\,\langle \mu_x \rangle \label{polarization}
\end{equation}
is taken as the expectation value of the atomic transition dipole moment $\mu_x$, where $n$ is the atomic density.

A self-consistent model is based on the numerical integration of Maxwell's equations\,(\ref{Faraday}) and\,(\ref{Ampere}), coupled via Eq.\,(\ref{polarization})
to the quantum dynamics. In the mean-field approximation employed here it is assumed that the density matrix of the ensemble is expressed as a product of density matrices of individual quantum emitters driven by a local EM field. In order to account for dipole-dipole interactions within a single grid cell we follow \cite{PhysRevA.47.1247} and introduce Lorentz-Lorenz correction for a local electric field term into quantum dynamics according to
\begin{equation}
\label{LL}
E_{x,\text{local}}=E_x+\frac{P_x}{3\varepsilon_0},
\end{equation}
where $E_x$ is the solution of Maxwell's equations (\ref{Faraday}), (\ref{Ampere}) and macroscopic polarization is evaluated according to Eq. (\ref{polarization}). 

\begin{figure}[t!]
\begin{center}
\includegraphics[width=0.48\textwidth]{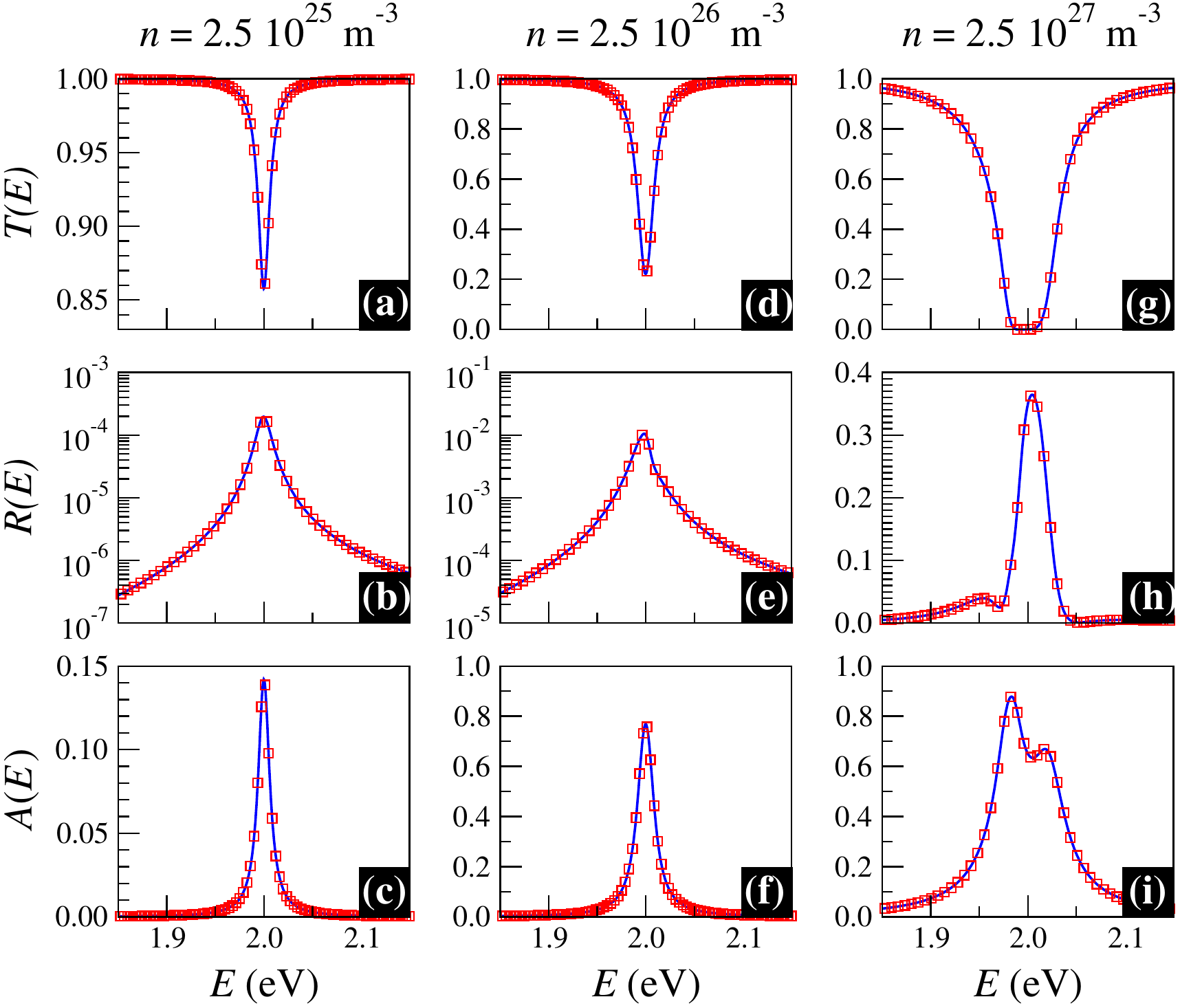}
\caption{\label{figure1} (Color online) Transmission $T(E)$ (panels a,d,g), reflection $R(E)$ (panels b,e,h) and absorption $A(E)$
(panels c,f,i) spectra of an atomic layer of thickness $\Delta z = 400$\,nm as a function of the incident photon energy $E$. The
atomic density is $n = 2.5 \times 10^{25}$\,m$^{-3}$ in the first column, $n = 2.5 \times 10^{26}$\,m$^{-3}$ in the second column,
and $n = 2.5 \times 10^{27}$\,m$^{-3}$ in the last column. The decay rate and pure dephasing rate are $\Gamma=10^{12}$\,s$^{-1}$
and $\gamma^*=10^{15}$\,s$^{-1}$, respectively. The atomic transition energy is $\hbar\omega_{\!B} = 2$\,eV and the transition dipole
moment is 2\,D. The solution of Maxwell-Liouville-von Neumann equations is shown as a blue solid line while the open red squares are from the
solution of our approximate non-Hermitian Schr\"odinger model.}
\end{center}
\end{figure}

The quantum dynamics is evaluated by computing the atomic dipole moment either using the single-atom density matrix\,(\ref{rho00})-(\ref{rho11})
or the single-atom wave packet\,(\ref{c0bis})-(\ref{c1bis}). To compare two approaches in the linear regime (\textit{i.e.} for $\rho_{11}\ll\rho_{00} \approx 1$,
$|c_1|^2 \ll |c_0|^2 \approx 1$), we calculate the transmission $T(E)$, reflection $R(E)$ and absorption $A(E)$ spectra of an atomic layer of thickness
$\Delta z = 400$\,nm as a function of the incident photon energy $E$ for an atomic transition energy of $E_B = \hbar\omega_{\!B} = 2$\,eV. The results
are shown in Fig.\,\ref{figure1}.

The transmission $T(E)$ and reflection $R(E)$ spectra are obtained from the normalized Poynting vector
\begin{equation}
 S = \frac{\left |\widetilde{E}_x \, \widetilde{H}_y \right |}{\left |\widetilde{E}_{x,\text{inc}} \, \widetilde{H}_{y,\text{inc}} \right |}
\end{equation}
at a specific location under and above the atomic layer, respectively, where $\widetilde{E}_x$, $\widetilde{H}_y$ and $\widetilde{E}_{x,\text{inc}}$,
$\widetilde{H}_{y,\text{inc}}$ are the Fourier components of the total and incident EM fields. The absorption spectrum is then simply obtained as
\begin{equation}
 A(E) = 1 - T(E) - R(E)\,.
\end{equation}

In Fig. \ref{figure1} the solution of Maxwell-Liouville-von Neumann equations is shown as a blue solid line while the open red squares are from the
solution of our approximate non-Hermitian Schr\"odinger model. One can notice a perfect agreement of the two methods irrespective of the atomic
density. These results clearly demonstrate that, in the linear regime where the atomic excitation probability remains small and varies linearly with
the incident field intensity, a simple wave packet propagation is sufficient to mimic the excitation and relaxation dynamics of an ensemble of quantum
emitters. Therefore in the weak field regime, the propagation of the full density matrix is unnecessary.

At low density, most of the incident radiation is simply transmitted and a small part of the incident energy is absorbed by the atomic ensemble at
energies close to the atomic transition energy of 2 eV. The absorption spectrum shows the conventional Lorentzian profile. At higher densities,
the absorption spectrum is strongly modified due to the appearance of collective excitation modes \cite{Sukharev:2011ib}, and a large part of
the incident energy is either absorbed or reflected from the atomic nano-layer at photon energies close to the transition energy.

\begin{figure}[t!]
\begin{center}
\includegraphics[width=0.41\textwidth]{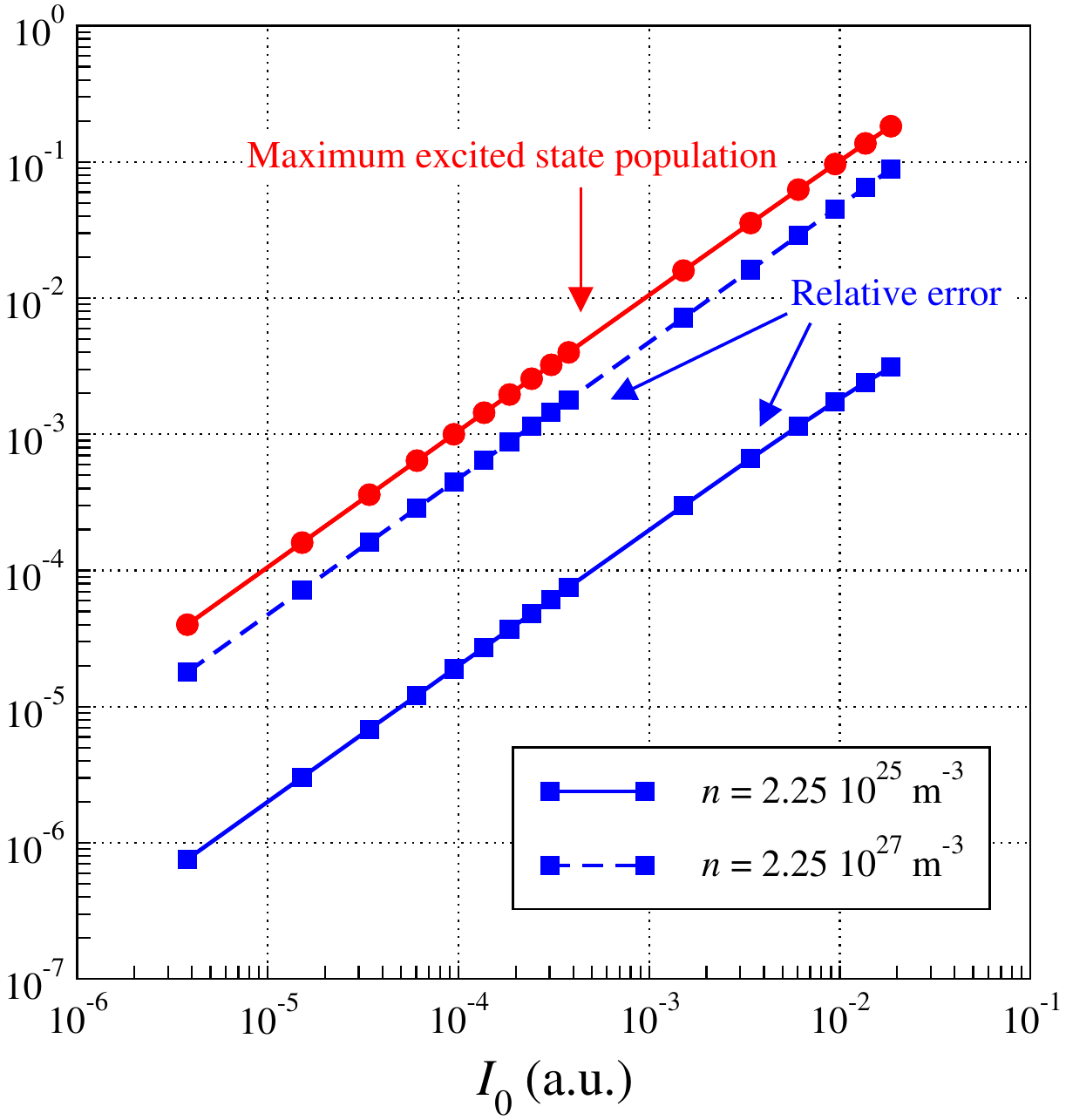}
\caption{\label{figure2} (Color online) Log-Log plot of the maximum excited state population (red solid line with circles) and the relative
error (blue lines with squares) in the calculation of the absorption spectrum $A(E)$ at the transition energy $E_B=\hbar\omega_{\!B}$ using the
Schr\"odinger approximation when compared to the solution of the full Liouville-von Neumann equation as a function of the incident field intensity
in atomic units. The blue solid line is for the lowest atomic density $n = 2.5 \times 10^{25}$\,m$^{-3}$ and the blue dashed line is for the
highest atomic density $n = 2.5 \times 10^{27}$\,m$^{-3}$. All other parameters are as in Fig. \ref{figure1}.}
\end{center}
\end{figure}

The blue lines with squares in Fig. \ref{figure2} show, as a function of the incident field intensity, the relative error $|A_S(E_{\!B})
- A_L(E_{\!B}) | \,/\, A_L(E_{\!B})$ obtained in the calculation of the absorption spectrum at the transition energy $E_{\!B}$ using the
Schr\"odinger approximation $A_S(E_{\!B})$ when compared to the solution of the full Liouville-von Neumann equation $A_L(E_{\!B})$.
The solid blue line is for the lowest atomic density $n = 2.5 \times 10^{25}$\,m$^{-3}$ and the dashed blue line is for the highest
density $n = 2.5 \times 10^{27}$\,m$^{-3}$.

One can see that irrespective of the atomic density the relative error of the non-Hermitian wave packet approximation scales linearly
with the field intensity. This is not surprising. Indeed, solving coupled Schr\"odinger equations (\ref{c0bis})-(\ref{c1bis}) is strictly
equivalent to solving the approximate Bloch equations (\ref{rhos00})-(\ref{rhos11}). The latter differ from the exact
Bloch equations (\ref{rho00})-(\ref{rho11}) by a term proportional to the excited state population $\rho_{11}$. The
maximum excited state population  $\rho_{11}^{\textrm{max}}$ is also shown in Fig. \ref{figure2} as a function of the field
intensity. It can be seen that it also varies linearly in the present weak field regime. We can conclude (from
Fig. \ref{figure2} and numerous calculations we have performed) that as long as the excited state population remains smaller
than 1\% our wave packet approximation can be used absolutely safely.

\begin{figure}[t!]
\begin{center}
\includegraphics[width=0.48\textwidth]{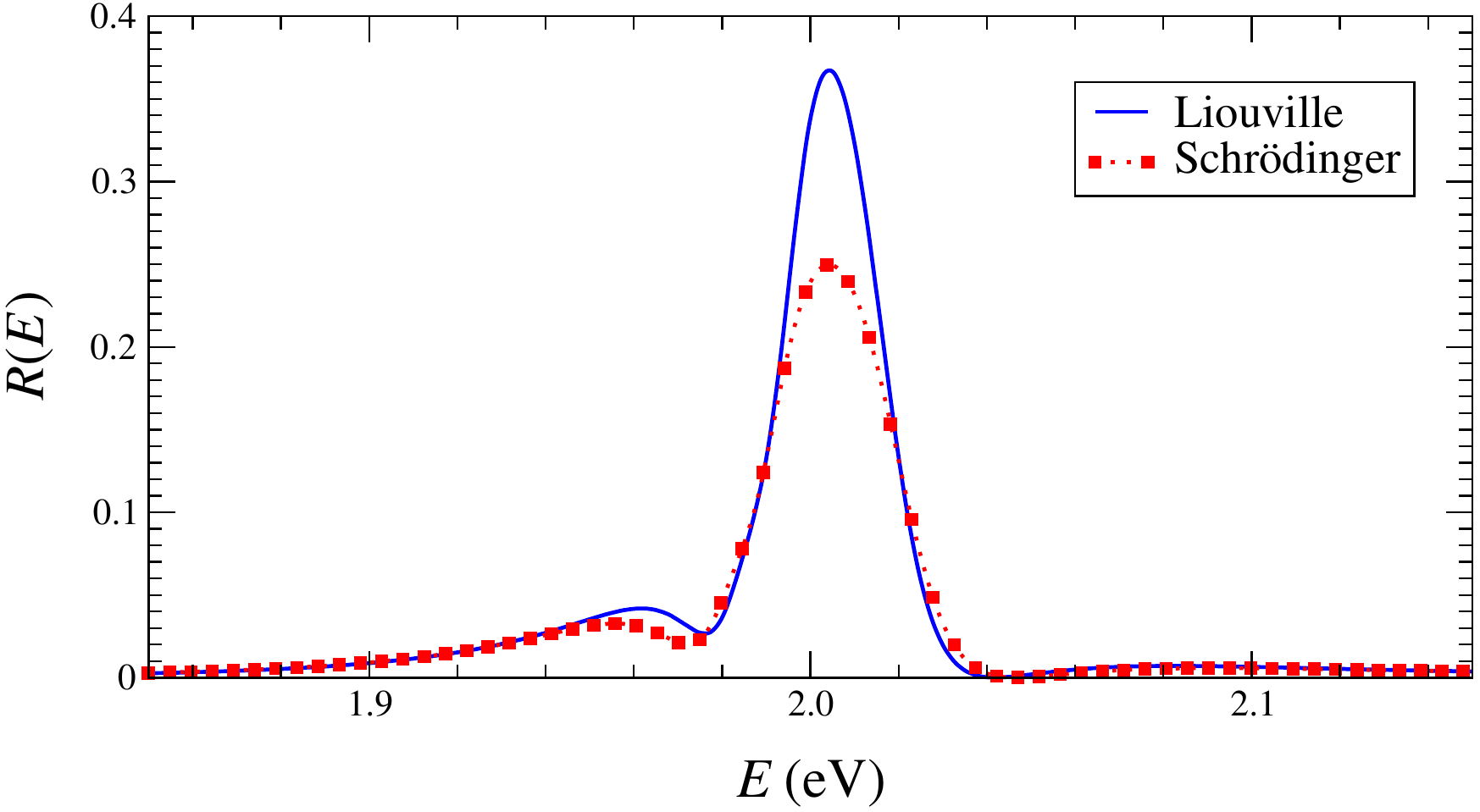}
\caption{\label{figure3} (Color online) Reflection probability $R(E)$ of an atomic layer of thickness $\Delta z = 400$\,nm as a function
of the incident photon energy $E$. The atomic density is $n = 2.5 \times 10^{27}$\,m$^{-3}$. The solution of Maxwell-Liouville-von-Neumann
equations is shown as a blue solid line while the dotted line with red squares is from the solution of our approximate non-Hermitian
Schr\"odinger model. The exciting field amplitude is chosen such that the maximum excited state population reaches 35\%. All other
parameters are as in Fig. \ref{figure1}.}
\end{center}
\end{figure}

As shown in Fig. \ref{figure3}, we could verify that the quality of the calculated spectra is still rather good when the
excited state population reaches 35\%. This figure shows the reflection spectra calculated using the ``exact'' Liouville-von-Neumann
equations and using our approximate Schr\"odinger model for a relatively high laser field amplitude, chosen such that
the maximum excited state population reaches 35\%. It is only when the excited state population approaches 50\% that one can observe
a very sudden failure of the present Schr\"odinger model, as could be expected from the divergence of the time-dependent gain and decay rates
$\gamma_0(t)$ (\ref{g0}) and $\gamma_1(t)$ (\ref{g1}) when $|c_0(t)|^2=|c_1(t)|^2$. This example shows that the present non-Hermitian
Schr\"odinger approximation still holds in the case of relatively large couplings.

\begin{figure}[t!]
\begin{center}
\includegraphics[width=0.48\textwidth]{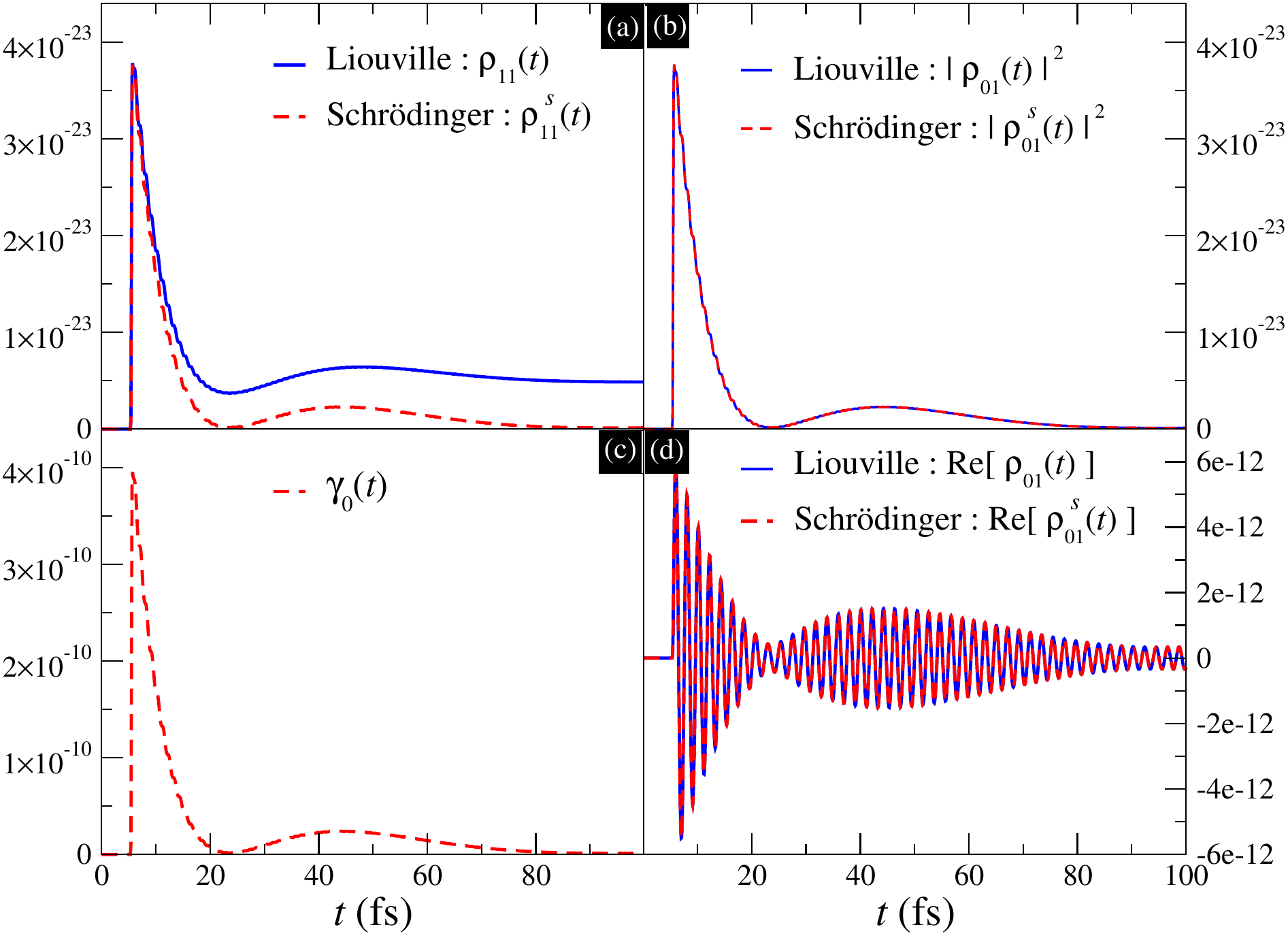}
\caption{\label{figure4} (Color online) Population dynamics: (a) excited state population as a function of time, (b) squared modulus
of the system's coherence as a function of time, (c) effective ground state gain rate $\gamma_0(t)$ as a function of time, (d) real
part of the system's coherence as a function of time. The results obtained from the solution of Liouville-von-Neumann equations are
shown with a blue solid line while the results obtained from the non-Hermitian Schr\"odinger approach are shown using red dashed lines.
The atomic density is $n = 2.5 \times 10^{27}$\,m$^{-3}$. All other parameters are as in Fig. \ref{figure1}.}
\end{center}
\end{figure}

The time dependence of the excited state population $\rho_{11}(t)$ and of the coherence $\rho_{01}(t)$ is illustrated in panels (a), (b) and
(d) of Fig. \ref{figure4}. These results were obtained with the same parameters as in Fig. \ref{figure1}, with an atomic density of
$n = 2.5 \times 10^{27}$\,m$^{-3}$. One can see from the panels (b) and (d) of this figure, showing the square modulus and the real part
of the coherence $\rho_{01}(t)$, that our non-Hermitian Schr\"odinger model reproduces quite accurately the coherence dynamics of the system. On the other
hand, as shown in panel (a), this is obtained at the cost of a poor description of the excited state population dynamics. Indeed, to
describe correctly the coherence of the system one is led to overestimate the excited state decay rate. However, as seen in Fig. \ref{figure1},
this overestimation of the decay rate does not have any impact on the accuracy of the calculated absorption, reflection and transmission spectra
when the excited state population remains small compared to the ground state population. Finally, panel (c) of Fig. \ref{figure4} shows the
time dependence of the gain coefficient $\gamma_0(t)$. This gain rate basically follows the evolution of the excited state population, as
it could already be inferred from Eq. (\ref{g0}). The decay rate $\gamma_1(t)$ is not shown in this figure since it is essentially constant and
equal to $(2\gamma^*+\Gamma)$ (see Eq. (\ref{g1})) in the present situation with $|c_1(t)|^2 \ll |c_0(t)|^2$.

\subsection{Generalization to multi-level systems}
\label{sec:multilevels}
In order to generalize our Schr\"odinger-type approximation of the excitation and dissipation dynamics of an ensemble of quantum emitters
to a multi-level system we now consider the case of neutral diatomic molecules. More specifically, we have chosen the particular case
of the ground and first excited electronic states of the Li$_2$ molecule and we follow the electronic dynamics and the nuclear motion by
expanding the total molecular wave function $\Psi(\mathbi{r}_{\!e\,},\mathbi{R},t)$ using the Born-Oppenheimer expansion
\begin{equation}
\Psi(\mathbi{r}_{\!e\,},\mathbi{R},t) =
\chi_g(\mathbi{R},t) \Phi_g(\mathbi{r}_{\!e\,}|R) + \chi_e(\mathbi{R},t) \Phi_e(\mathbi{r}_{\!e\,}|R)
\end{equation}
where $\Phi_g(\mathbi{r}_{\!e\,}|R)$ and $\Phi_e(\mathbi{r}_{\!e\,}|R)$ denote the electronic wave functions associated with the ground
$X\rm\left(^{1}\Sigma_{g}^{+}\right)$ and first excited $A\rm\left(^{1}\Sigma_{u}^{+}\right)$ electronic states of Li$_2$, respectively.
The electron coordinates are denoted by the vector $\mathbi{r}_{\!e\,}$, and the vector $\mathbi{R} \equiv (R,\hat{R})$ represents the
internuclear vector. 

We now separate the global electronic coordinate $\mathbi{r}_{\!e\,}$ of all electrons into the coordinate $\mathbi{r}_{\!c\,}$ of the
core electrons and the coordinate $\mathbi{r}$ of the active electron \cite{Charron:2006my}. The ground $X\rm\left(^{1}\Sigma_{g}^{+}\right)$
electronic state is considered as a 2s$\sigma$ state, and the electronic wave function $\Phi_g(\mathbi{r}_{\!e\,}|R)$ is expressed in
the molecular frame (Hund's case (b) representation) as the product
\begin{equation}
\label{eq:Phig}
\Phi_g(\mathbi{r}_{\!e\,}|R) = \phi_{g}(\mathbi{r}_{\!c\,}|R)\,R_X(r)\,Y_{00}(\hat{r})
\end{equation}
where $R_X(r)$ and $Y_{00}(\hat{r})$ are the radial and angular parts of the electronic wave function associated with the active
electron. Similarly, the 2p$\sigma$ excited state of $A\rm\left(^{1}\Sigma_{u}^{+}\right)$ symmetry is expressed in the molecular frame as
\begin{equation}
\label{eq:Phie}
\Phi_e(\mathbi{r}_{\!e\,}|R) = \phi_{e}(\mathbi{r}_{\!c\,},r|R)\,R_A(r)\,Y_{10}(\hat{r})\,.
\end{equation}

Due to the $\Sigma$ symmetry of both electronic states, the ro-vibrational time-dependent wave functions $\chi_g(\mathbi{R},t)$ and
$\chi_e(\mathbi{R},t)$ can be expanded on a limited set of normalized Wigner rotation matrices in order to take into account the
rotational degree of freedom, following \cite{PhysRevA.49.R641}
\begin{subequations}
 \begin{eqnarray}
  \chi_g(\mathbi{R},t) & = & \sum_{N,M} \chi^{g}_{N,M}(R,t) \, D^{N^{\,*}}_{M,0}(\hat{R})\\
  \chi_e(\mathbi{R},t) & = & \sum_{N,M} \chi^{e}_{N,M}(R,t) \, D^{N^{\,*}}_{M,0}(\hat{R})\,,
 \end{eqnarray}
\end{subequations}
where $N$ denotes the molecular rotational quantum number while $M$ denotes its projection on the electric field polarization axis $x$
of the laboratory frame.

Introducing these expansions in the time-dependent Schr\"odinger equation (\ref{Schrodinger}) describing the molecule-field interaction and
projecting onto the electronic and rotational basis functions yields, in the dipole approximation, the following set of coupled differential
equations for the nuclear wave packets $\chi^{g}_{N,M}(R,t)$ and $\chi^{e}_{N,M}(R,t)$
\begin{subequations}
\label{eq:tdse}
\begin{eqnarray}
i\hbar\frac{\partial}{\partial t}\chi^{g}_{N,M} & = & \hat{\cal H}^{g}_{N}(R)\,\chi^{g}_{N,M} - E_x(t) \mu_{AX}(R)\nonumber\\
                                                &   & \qquad\times \sum_{N',M'} {\cal M}^{N',M'}_{N,M} \chi^{e}_{N',M'}\label{eq:chig}\\
i\hbar\frac{\partial}{\partial t}\chi^{e}_{N,M} & = & \hat{\cal H}^{e}_{N}(R)\,\chi^{e}_{N,M} - E_x(t) \mu_{AX}(R)\nonumber\\
                                                &   & \qquad\times \sum_{N',M'} {\cal M}^{N',M'\,^*}_{N,M} \chi^{g}_{N',M'}\label{eq:chie}
\end{eqnarray}
\end{subequations}
where $\mu_{AX}(R) = \langle R_A | r | R_X \rangle_{\!r}$ is the electronic transition dipole. The ro-vibrational nuclear Hamiltonians
$\hat{\cal H}^{g/e}_{N}(R)$ are defined as
\begin{equation}
\hat{\cal H}^{g/e}_{N}(R) = -\frac{\hbar^2}{2\mu}\left[\frac{\partial^2}{\partial R^2}-\frac{N(N+1)}{R^2}\right] + V_{g/e}(R)
\end{equation}
where $\mu$ denotes the molecular reduced mass. The potential energy curves $V_{g}(R)$ and $V_{e}(R)$ associated with the ground and
first excited electronic states of the molecule are taken from Ref. \cite{SchmidtMink1985263} and the matrix elements ${\cal M}^{N',M'}_{N,M}$
which couple the nuclear wave packets evolving on these electronic potential curves can be written using $3j$-symbols as
\begin{eqnarray}
\label{eq:coupling}
{\cal M}^{N',M'}_{N,M} & = & (-1)^M \frac{(2N+1)}{\sqrt{4\pi}} \sum_{N',M'} \left( \begin{array}{ccc} N' & 1 & N\\ 0  & 0 &  0 \end{array} \right)\nonumber\\
                       &   &                 \qquad\qquad\qquad\quad \times \left( \begin{array}{ccc} N' & 1 & N\\ M' & 0 & -M \end{array} \right)
\end{eqnarray}

We assume that the molecules are prepared at time $t=0$ in the ro-vibrational level $v=0$ and $N=0$ of the ground electronic state.
In weak linearly polarized fields and except for a phase factor this ground state component of the molecular wave function remains
unaffected and the excited state component is limited to $N=1$ and $M=0$. The ground and excited nuclear wave packets are thus
finally expanded in terms of ro-vibrational eigenstates as
\begin{subequations}
 \begin{eqnarray}
  \chi^{g}_{0,0}(R,t) & = &          c_g(t)\,\varphi^{g}_{0,0}(R)\\
  \chi^{e}_{1,0}(R,t) & = & \sum_{v} c_v(t)\,\varphi^{e}_{v,1}(R)
 \end{eqnarray}
\end{subequations}
where $c_g(t)$ and $c_v(t)$ are time-dependent complex coefficients. $\varphi^{g/e}_{v,N}(R)$ denote here the bound ro-vibrational eigenstates
of the $g/e$ electronic potentials \cite{Charron:3922}. It is not necessary here to take into account the dissociative nuclear eigenstates
associated with the excited potential since their coupling with the ground vibrational level of the ground electronic state is negligible.

We thus arrive at a multi-level system which is very similar to the atomic case described in sections \ref{sec:tls} and \ref{sec:nhtlwpa},
except that the initial ground state is now coupled with a large set of excited levels. For convenience and for an easy comparison with the
atomic case, we will label the ground state as state number 0 and the excited states as states number $j \geqslant 1$. Our reference
calculations will be based on the numerical solutions of the corresponding Liouville-von Neumann equations
\begin{subequations}
 \begin{eqnarray}
  \dot{\rho}_{00} & = & \sum_{j \geqslant 1} \big[ i\Omega_j(t)\left(\rho_{0j}-\rho_{j0}\right) + \Gamma\rho_{jj} \big] \label{rho00new}\\
  \dot{\rho}_{0j} & = & i\Omega_j(t)\left(\rho_{00}-\rho_{jj}\right)\nonumber\\
                  &   & \qquad + \left[i(\omega_j-\omega_0) - \frac{2\gamma^*+\Gamma}{2}\right]\rho_{0j} \label{rho0j}\\
  \dot{\rho}_{j0} & = & i\Omega_j(t)\left(\rho_{jj}-\rho_{00}\right)\nonumber\\
                  &   & \qquad - \left[i(\omega_j-\omega_0) + \frac{2\gamma^*+\Gamma}{2}\right]\rho_{0j} \label{rhoj0}\\
  \dot{\rho}_{jj} & = & i\Omega_j(t)\left(\rho_{j0}-\rho_{0j}\right) - \Gamma\rho_{jj} \label{rhojj}
 \end{eqnarray}
\end{subequations}
where $\hbar\omega_j$ is the total energy of the excited state $j$ and where $\Omega_j(t)$ denotes the instantaneous Rabi frequency
associated with the molecule-field interaction as defined in Eqs.\,(\ref{eq:chig}) and\,(\ref{eq:chie}). $\gamma^{*}$ and $\Gamma$
denote the pure dephasing rate and the relaxation rate associated with the excited states, respectively.

In comparison with this ``exact'' model, our Schr\"o\-din\-ger-type approximation will be based on the numerical solutions of the coupled
equations for the time-dependent expansion coefficients
\begin{subequations}
 \begin{eqnarray}
  i\,\dot{c}_0 & = & \left[ \omega_0 + i\frac{\gamma_0(t)}{2} \right] c_0 + \sum_{j \geqslant 1} \Omega_j(t)\,c_j \label{c0ter} \\
  i\,\dot{c}_j & = & \Omega_j(t)\,c_0 + \left[ \omega_j - i\frac{\gamma_j(t)}{2} \right] c_j\,, \label{c1ter}
 \end{eqnarray}
\end{subequations}
where the time-dependent gain and decay rates $\gamma_0(t)$ and $\gamma_j(t)$ are now defined as
\begin{eqnarray}
 \gamma_0(t) & = & \displaystyle\frac{(2\gamma^{*}+\Gamma)\sum_{j \geqslant 1}|c_j(t)|^2}{|c_0(t)|^2-\sum_{j \geqslant 1}|c_j(t)|^2} \label{g0c}\\
 \gamma_j(t) & = & \displaystyle\frac{(2\gamma^{*}+\Gamma)                    |c_0(t)|^2}{|c_0(t)|^2-\sum_{j \geqslant 1}|c_j(t)|^2} \label{g1c}
\end{eqnarray}

One can show that with such a definition of $\gamma_0(t)$ and $\gamma_j(t)$, Eqs.\,(\ref{c0ter}) and\,(\ref{c1ter}) are strictly equivalent to
Eqs.\,(\ref{rho00new})-(\ref{rhojj}) in the limit of weak couplings. In the next section we demonstrate that our Schr\"odinger-type model
accurately reproduces the excitation and dissipation dynamics of multi-level quantum systems in the limit of weak couplings, \textit{i.e.} for
$\sum\limits_{j \geqslant 1} |c_j|^2 \ll |c_0|^2 \approx 1$.

\subsection{Application to a uniform nano-layer of molecules}
\label{sec:molecular_layer}

We consider a simplified one-dimensional system similar to the one discussed in case of two-level atoms. A uniform infinite layer with a thickness
of $\Delta z = 400$ nm comprised of Li$_2$ molecules is exposed to incident linearly polarized radiation. The incident field propagates in the positive
$z$-direction. We calculate the absorption spectrum $A(E)$ of this molecular layer just as we did in section \ref{sec:aunla} for atoms.

\begin{figure}[t!]
\begin{center}
\includegraphics[width=0.48\textwidth]{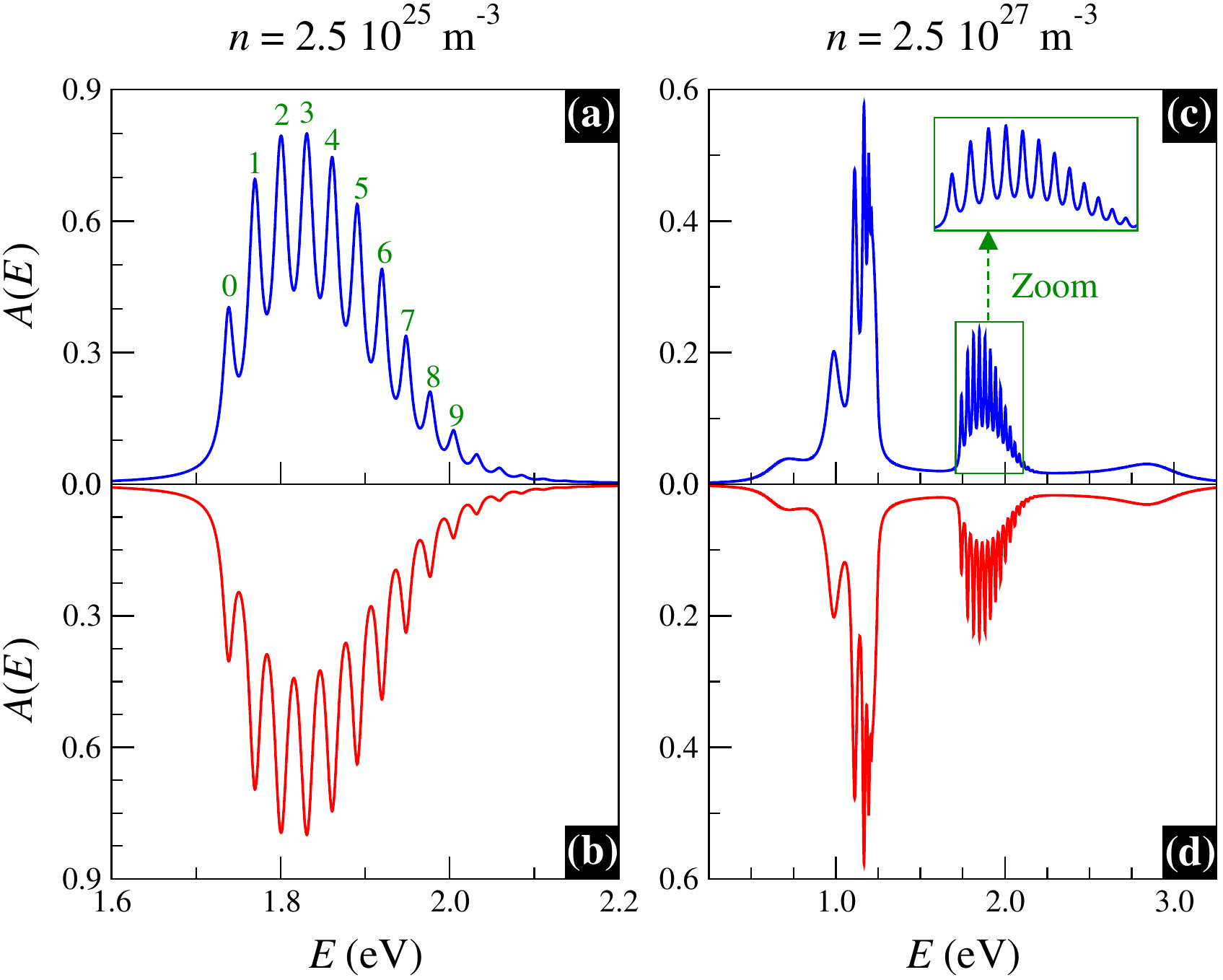}
\caption{\label{figure5} (Color online) Absorption spectra $A(E)$ of a Li$_2$ molecular layer of thickness $\Delta z = 400$\,nm as a
function of the incident photon energy $E$. The molecular density is $n = 2.5 \times 10^{25}$\,m$^{-3}$ in the left column (panels
(a) and (b)) and $n = 2.5 \times 10^{27}$\,m$^{-3}$ in the right column (panels (c) and (d)). The solutions of Maxwell-Liouville-von Neumann
equations are shown as blue solid lines in the first raw (panels (a) and (c)) while the red solid lines (inverted spectra, panels (b)
and (d)) are from the solutions of our approximate non-Hermitian Schr\"odinger model. All other parameters are as in Fig. \ref{figure1}.}
\end{center}
\end{figure}

To ascertain the validity of the proposed Schr\"odinger-type approximation, absorption spectra are represented in Fig. \ref{figure5}
as a function of the incident photon energy $E$, in the linear regime, for two different molecular densities: $n = 2.5 \times 10^{25}$\,m$^{-3}$
in the left column (panels (a) and (b)) and $n = 2.5 \times 10^{27}$\,m$^{-3}$ in the right column (panels (c) and (d)). The solutions obtained
via integrating Maxwell-Liouville-von Neumann equations are shown in the upper panels (a) and (c) as blue solid lines while the ones obtained from our approximate
non-Hermitian Schr\"odinger model are shown in red in the lower panels (b) and (d) (inverted spectra).

As in the case of two-level atoms, one can notice a perfect agreement of the two methods irrespective of the molecular density. This shows that,
in the linear regime where the molecular excitation probability remains small and varies linearly with the incident field intensity, a simple
wave packet propagation is sufficient to mimic the excitation, dissipation and decoherence dynamics of an ensemble of multi-level quantum emitters.
The propagation of the full density matrix is then, again, unnecessary.

It is important to note that at low density, we observe a series of overlapping vibrational resonances which reflects the vibrational structure of the excited molecular
potential and which follows the Franck-Condon principle \cite{TF9262100536,PhysRev.28.1182}. The green labels seen in panel (a) of Fig. \ref{figure5} indicate the
excited state vibrational level responsible for the observed resonance.

\begin{figure}[t!]
\begin{center}
\includegraphics[width=0.41\textwidth]{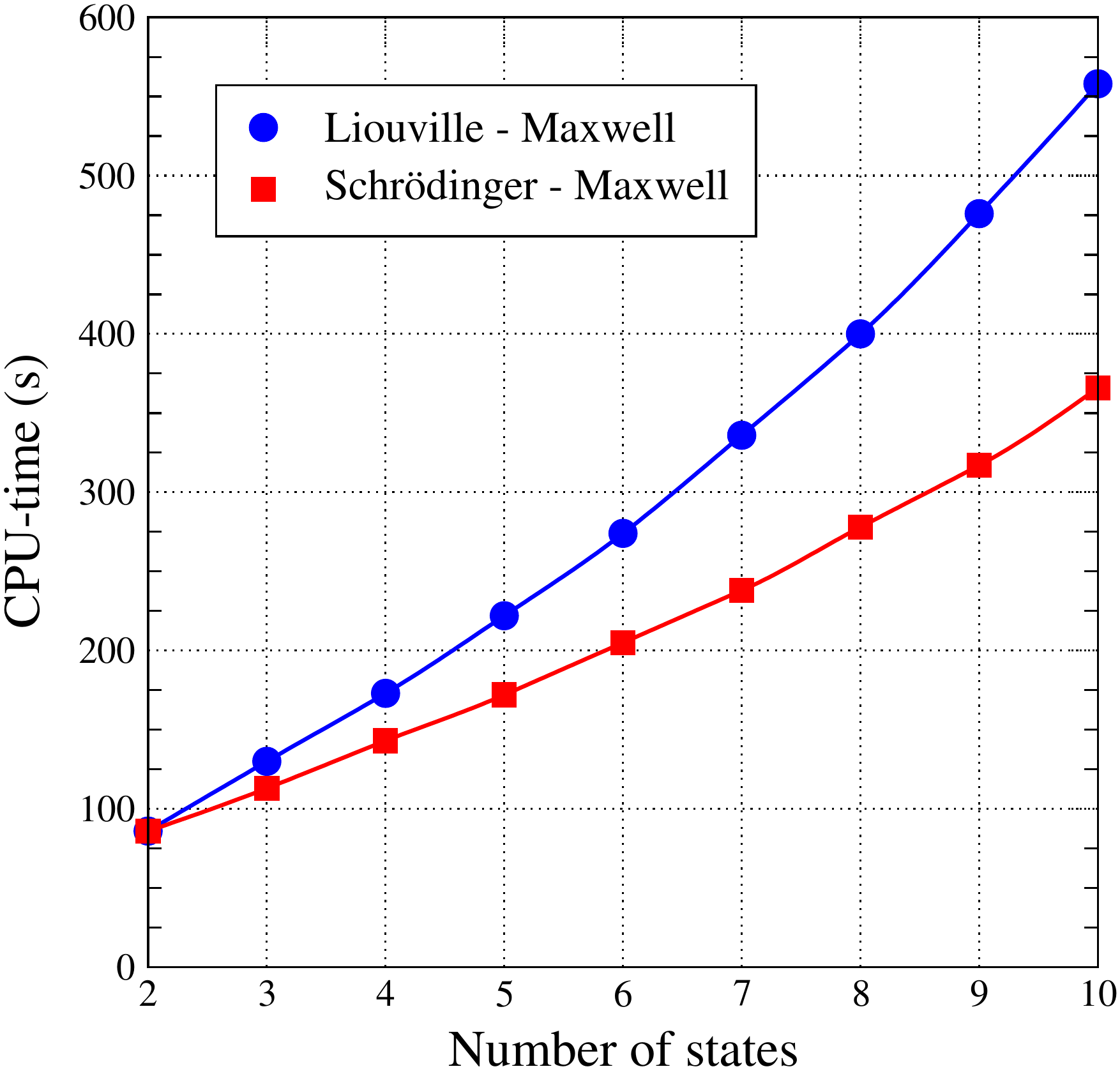}
\caption{\label{figure6} (Color online) Computation process time necessary on a Intel Xeon E5-1650 processor for the calculation of the absorption spectrum
of a Li$_2$ molecular nano-layer of thickness $\Delta z = 400$\,nm as a function of the number of quantum levels included in the calculation. The blue
line with circles is for the solution obtained from Maxwell-Liouville-von Neumann equations while the red line with squares is for our proposed Schr\"odinger-type
approximation. The spatial grid has a total size of 2560\,nm with a spatial step of 1\,nm. The time propagation is performed on a temporal grid of total
size 1.7\,ps with a time step of 1.7\,as.}
\end{center}
\end{figure}

At higher density, the absorption spectrum is strongly distorted and one observes, just like in the atomic case \cite{Sukharev:2011ib}, the appearance of collective
excitation modes, the difference being that a vibrational structure may still be present in some of these collective molecular excitation modes. A detailed analysis
of the physics underlying the appearance of these intriguing molecular collective modes will be presented in another paper. For high densities, we could also observe
that the radiation field is not transmitted anymore through the molecular layer in the absorption window $1.2\,\textrm{eV} \leqslant E \leqslant 2.5\,\textrm{eV}$
of the molecule. Within this window, the field is either absorbed or reflected. The small inset seen in panel (c) of Fig. \ref{figure5} shows a magnification of the
absorption spectrum in the energy range $1.7\,\textrm{eV} \leqslant E \leqslant 2.1\,\textrm{eV}$ corresponding to the ``normal'' low-density absorption spectrum. One
can see in this inset, and by comparing with panel (a) of Fig. \ref{figure5}, that the absorption spectrum is not strongly modified at high molecular density in this
energy region.

Figure\,\ref{figure6} finally shows the computation time necessary for the calculation of these absorption spectra as a function of the number
of quantum states introduced in this multi-level model for both the Maxwell-Liouville-von Neumann (blue line with circles) and Maxwell-Schr\"odinger
(red line with squares) approaches. One can observe a substantial difference in computation times which can prove of crucial importance when one
has to deal with realistic three-dimensional systems. This origin of the observed gain in computation time relies on the necessity of
propagating a single wave function instead of a full density matrix.

\section{Summary and conclusions}
\label{sec:summary}

We proposed a new and simple non-Hermitian approximation of Bloch optical equations where one propagates the wave function of the quantum system instead of the
complete density matrix. Our method provides an accurate, complete description of the excitation, relaxation and decoherence dynamics of single as well as ensembles
of coupled quantum emitters (atoms or molecules) in weak EM fields, taking into account collective effects and dephasing. We demonstrated the applicability of the
method by computing optical properties of thin layers comprised of two-level atoms and diatomic molecules. It was shown that, in the limit of weak incident fields,
the dynamics of interacting quantum emitters can be successfully described by our set of approximated equations, which result in a substantial gain both in computation
time and computer memory requirements. These calculations also reveal some intriguing new collective molecular excitation modes which will be presented in detail in
another publication. The proposed approach was demonstrated to provide a substantial increase in numerical efficiency for self-consistent simulations.

\section{Acknowledgements}

E.C. would like to acknowledge useful and stimulating discussions with O. Atabek and A. Keller from Universit\'e Paris-Sud (Orsay) and with E. Shapiro from the
University of British Columbia (Canada). M.S. is grateful to
the Universit\'e Paris-Sud (Orsay) for the financial support through an invited Professor position in 2011. E.C. acknowledges supports from ANR (contract
Attowave ANR-09-BLAN-0031-01), and from the EU (Project ITN-2010-264951, CORINF). We also acknowledge the use of the computing facility cluster GMPCS of the LUMAT
federation (FR LUMAT 2764).

\end{document}